\documentclass[iop]{emulateapj-rtx4} % read by: ACROREAD
\shortauthors{Sekanina} %, Skiff, \& Green}
\shorttitle{On Comet Wirtanen (C/1956 F1)}
\slugcomment{Version \today }

\begin{document}
\title{On Elusive Observations and a Sly Companion of Comet Wirtanen
 (C/1956 F1)}
\author{Zdenek Sekanina}
\affil{La Canada Flintridge, California 91011, U.S.A.}
\email{ZdenSek@gmail.com.}

\begin{abstract} % maximum length = 1920 characters
Noting that the extensive astrometric observations of the double comet
Wirtanen (C/1956~F1)~made by E.\ Roemer have never been published, I replicate
the contents of a fortuitously discovered copy~of her measurement records of
the companion's offsets from the main mass in 1957--1959~and~use~with~such
data by others to refine the fragmentation solution.  The sublimation-driven
nongravitational acceleration is shown to essentially control the
companion's motion in the orbital plane.  The~\mbox{fragmentation}
parameters derived by the author in 1978 have now been improved
and strong disagreement~with~the~independent results by Roemer is noted.
The revised model is employed to predict the positions of the companion
on the plates exposed by Roemer on 25 September 1960, which she reported
to~show~the principal nucleus but not the companion.  At my request,
these plates have now been scanned and processed at the Lowell
Observatory, and the companion is found to be located at the predicted
position.  The images of the main mass and the companion on one
of the two plates are displayed.
\end{abstract}
\section{Introduction} %%% Sec. 1
\vspace{-0.08cm}
Comet Wirtanen, listed nowadays in the catalogues as C/1956 F1, was originally
designated as 1956c~=~1957~VI.  Discovered by C.~A.~Wirtanen with the
51-cm Carnegie astrograph of the Lick Observatory on 16~March~1956, the
comet was an extraordinary object.  Having arrived from the Oort Cloud,
it reached perihelion at 4.45~AU and was observed over a period of
4.5~years, for the last time at a heliocentric distance of 9.43~AU, the
second largest up to that time.  The comet possessed~a~\mbox{narrow},
parallel-sided tail, typical for some distant comets, but the most
salient feature was a second nucleus, detected first by Van Biesbroeck
(1961) with the 208-cm f/3.9 reflector of the McDonald Observatory at
the beginning of May 1957, four months prior to perihelion, and then
observed steadily over at least 28~months.  I show~\mbox{below} that the
companion was in fact seen for as long as the primary, the overall
arc amounting to 41~months.

Because in the months after its discovery the companion was 2.5 to 3 magnitudes
fainter than the main nucleus, it was detected only with large instruments.
Next to Van Biesbroeck, the companion's observations were reported by a
team of astronomers at the Lick Observatory (Jeffers \& Klemola 1958;
Jeffers \& Gibson 1960) as well as by Roemer (1962, 1963) at the U.S.\
Naval Observatory's Flagstaff Station.  Whereas Van Biesbroeck and the Lick
observers published the astrometric positions of both the principal
nucleus and the companion, Roemer's accounts were limited to reports
of descriptive nature.

As time went on, the companion appears~to~have increasingly been
shedding some of its mass as microscopic dust and, as a result, the
magnitude difference between the two fragments was gradually diminishing,
averaging $\sim$1.5~magnitudes in 1958 and 1~magnitude~in~1959.
Accordingly, the companion's brightness remained within the reach of
the large telescopes even as the primary was progressively fading as
the comet was receding from the Sun.  At Lick the comet was last
observed in July 1959, at McDonald Van Biesbroeck detected it as late
as September 1959.  In either case the companion was still visible.
On the other hand, Roemer (1960) extended her observing run~for another
year, into September 1960, noting however that the comet  ``{\it is
now so faint that the second nucleus cannot be photographed even if
it is still present.\/}''  I will return to this topic in Section~5.

\section{Roemer's Investigation of the Splitting of Comet Wirtanen} % Sect. 2
% \section{Splitting of Comet Wirtanen Investigated\\By the Late Elizabeth
% Roemer}
%
Roemer (1963) reported that about 60~plates showing the two nuclei
were exposed at Flagstaff, a remarkable achievement in comparison with
12~positions secured at Lick and 19 images provided by Van Biesbroeck,
many regrettably of lower resolution, taken with the 61-cm f/4
reflector at Yerkes.  Roemer was unquestionably in the possession of
a superb dataset, but her 1962 and 1963~reports suggested that the
results on the nuclear duplicity were limited to a plot of the
companion's projected distance from the main nucleus as a function of time
fitted by a straight line.  The distance was corrected for the effects of
the geocentric distance, but otherwise the varying Earth-Sun-comet
geometry was not accounted for.

This somewhat underwhelming approach resulted in the instant of splitting
near 1~January 1957,~\mbox{determined} by extrapolating to the time of
zero separation distance, and in the projected separation velocity of
$\sim$1.5~m~s$^{-1}\!$, derived from the slope of the data fit.  Roemer
considered this latter quantity, which obviously was a lower~limit~of the
genuine separation velocity (in the framework of the adopted interpretation
scheme), as an upper limit of the velocity of escape from the nucleus of
comet Wirtanen to estimate its mass.  Her result, equaling~10$^{17}$\,grams,
is at best~an order-of-magnitude guess that should be taken with a great
deal of caution given that the breakup of a comet's nucleus is a complex
physical process and the motion of a fragment is affected by forces that
Roemer ignored.  The reader is referred to Appendix A for more
details.\\

\section{Scientific Interest in Comet Wirtanen and Its Fragmentation}
% Section 3
\vspace{-0.07cm}
In the mid-1960s Roemer and her collaborators~published an impressive
``trilogy'' of papers (Roemer~1965; Roemer \& Lloyd 1966; and Roemer et
al.\ 1966),~which contained more than 2000~astrometric observations~of
comets, unusual asteroids, and satellites.  It was~a~compendium of her
astrometric\,(and~photometric)\,work~made during the nearly 10~years of
observing at Flagstaff.  And while Roemer maintained that {\it
``observations of almost every comet to appear in the interval are
included,''} not a single observation of comet Wirtanen was, with no
explanation.  The attentive reader that registered this peculiarity
was inevitably left with an impression that the comet was deliberately
singled out for exclusion.

Given Roemer's (1962) statement that Wirtanen was an ``{\it exceptionally
interesting comet for several reasons\/}'', this plain case of data-release
suppression was incomprehensible.  Moreover, inspection of Roemer's {\it
complete\/} list of publications shows that her astrometric positions of
comet Wirtanen were {\it never\/} published.  The lack of contribution
of the astrometry from her as a leader among comet observers of the time
manifested itself in practice.  In particular, it adversely affected the
orbit determination of comet Wirtanen, which was highly desirable for
confirming the object's Oort Cloud membership.  When Marsden et al.\
(1978) went ahead with the computation, a pitiful total of 37~astrometric
observations was available that satisfied the high standards of accuracy
required.  Similarly restricted was the pool of data for studies of the
tail orientation, which provided information about the comet's onset of
activity on the way to perihelion (e.g., Sekanina 1975).

The most critical was the issue of Wirtanen as a split comet.
Both theoretical and observational considerations led the author to
fundamentally rethink the problem of cometary fragmentation.  The primary
theoretical argument was anchored in the basic idea of Whipple's (1950)
icy conglomerate model:\ If the sublimation-driven nongravitational
forces affected the motions of cometary nuclei, they must have influenced
the relative motions of cometary fragments as well.

The second argument was based on the properties of the orbital
motion of a comet's companion relative to the main mass.  Shortly
after the parent's breakup, when the fragments are seen close to one
another (say, several arcsec apart), the companion is receding from
the primary essentially along the prolonged radius vector.  On the
other hand, in wide double comets, long after separation, the
companion trails the primary along the orbit.  This kind of behavior
is known to be displayed by dust particles in a tail, as they move
across it from its sharp boundary nearer the nucleus to the diffuse
boundary far from the nucleus, being subjected to a repulsive force
--- the solar radiation pressure.  The sublimation of gases
predominantly from the sunlit hemisphere and preferably in the
sunward direction generates, in the opposite, antisolar direction,
a momentum that triggers an acceleration of any fragment.  However,
since the acceleration varies as the cross-sectional area and
inversely as the mass, the smaller the fragment the higher is the
acceleration in the antisolar direction.  Hence, the companion is
expected to move relative to the primary fundamentally in the same
manner as a dust-tail particle relative to the nucleus.

The third argument involved the traditional treatment of the orbital
motion of a comet's fragment as a consequence of its separation velocity.
If the nongravitational acceleration were ignored, the computer code
would inevitably attribute its effect to the impulse acquired by the
companion at separation, thereby artificially augmenting the magnitude
of the separation velocity.  This result was distinctly perceived in
Stefanik's (1966) list of 13~split comets.  Disregarding the
nongravitational effect, he derived separation velocities of
up to nearly 40~m~s$^{-1}$ and averaging $\sim$15~m~s$^{-1}$!  Comet
Wirtanen appeared {\vspace{0.02cm}}at the lower end of the range:\
at 2~m~s$^{-1}$ the velocity was in line with Roemer's value, but
--- as shown below --- still too high.

The net outcome of these considerations was the formulation of
a conceptually new model for the motions of the split comets (Sekanina
1977), which next to the time of breakup introduced the magnitude of
the sublimation-driven nongravitational acceleration as a new basic
parameter.  The variation of this acceleration with heliocentric distance
was approximated by an inverse square power law and its magnitude was
expressed in units of 10$^{-5}$\,the solar gravitational acceleration.

The model was applied to nearly two dozen companions of more than a
dozen split comets and provided rather satisfactory results in almost
all cases.  The most prominent systematic deviations between the model
and observations were exhibited by none other than comet Wirtanen, for
which only 25~data points were available in the absence of Roemer's
observations.  In general,~the lowest accelerations obtained were a
few units, referring to the most sizable companions.  For comparison
with~the units used{\vspace{-0.03cm}} by Marsden et al.\ (1973) in
their orbital computations, \mbox{$1\:\!\!\times\!10^{-5}$} the
solar gravitational acceleration equaled
\mbox{$0.3\:\!\!\times\!10^{-8}$}\,AU~day$^{-2}$ at 1~AU from the Sun.

The failure of the model in the case of comet~\mbox{Wirtanen} was
prompted by the companion's motion in the out-of-plane direction
(Section~4), which could not be fitted by an acceleration assumed to
point radially away from the Sun and therefore in the orbital plane.
The model was modified by adding three orthogonal components~of the
separation velocity --- in the radial, transverse, and normal directions
--- that made it a five-parameter model.  The software was written so that
any combination of one to five parameters could be solved for, a choice of
31~different versions of the model.  This was necessary because sometimes
there were strong correlations between some parameters (rather often, for
example, between the time of splitting and the radial component of the
separation velocity) and the convergence of the least-squares routine
could then be slow or unachievable.
% \vspace{0.1cm}

\section{Unforeseeable Developments and\\New Computations} % Section 4
%
% \vspace{0.15cm}
Application of the expanded model to eight extensively observed companions
of five split comets (Sekanina 1978) offered~much improved solutions,
including an excellent fit to 22 observations of comet Wirtanen.  However,
the results were strongly contradicting Roemer's findings, as is apparent
from the following.

\begin{table}[t]
\vspace{0.14cm} % 0.13
\hspace{-0.1cm}
\centerline{
\scalebox{0.983}{  % 0.985
\includegraphics{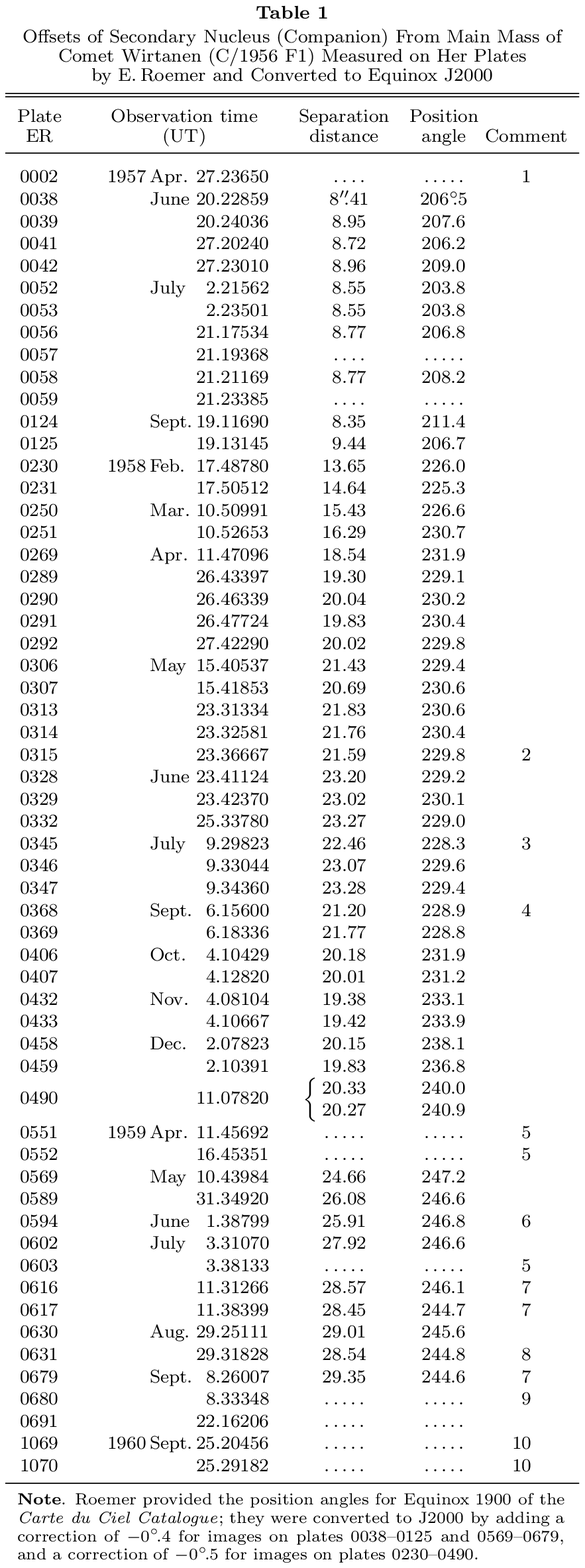}}}   %  TABLE 1
\end{table}

One, the five-parameter solution, fitting the 22~data with a mean
residual of $\pm$0$^{\prime\prime\!}$.56, indicated that the parent
nucleus broke up not around 1~January~1957, but in mid-September 1954
(with a formal uncertainty of about $\pm$2~months), that is, more than
two years earlier than determined by Roemer.  The implication was that
discovered in March 1956 was the primary fragment, not the original nucleus;
the companion was unobservable in 1956 because of its presumed faintness
and distance from the primary of merely \mbox{2--3$^{\prime\prime}$}.
The companion's differential nongravitational acceleration was found
to amount to some 7~units of 10$^{-5}$\,the solar gravitational
acceleration (with an uncertainty of about $\pm$10~percent), equivalent
to approximately \mbox{$2 \:\!\!\times \!10^{-8}$\,AU day$^{-2}$} at
1~AU from the Sun.  The companion appears to have been a fairly massive
chunk of material, probably hundreds of meters across.  The {\it
total\/} velocity of separation came out to be as low as
0.26\,$\pm$\,0.02~m~s$^{-1}\!$, very nearly perpendicular to the orbital
plane, and a factor of about six smaller than Roemer's value, which
referred besides to the component projected onto the plane of the sky.

Two, the model's versatility allowed me to test whether a solution using the
separation velocity alone was competitive in data fitting.  Optimizing the
four parameters on the condition of {\it no nongravitational acceleration\/},
a total separation velocity came out to be 0.44\,$\pm$\,0.03~m~s$^{-1}\!$,
about 70~percent higher, and the splitting took place in January 1955
with a 1$\sigma$ uncertainty of nearly $\pm$4~months.  In comparison to the
five-parameter solution, the quality of fit unfortunately deteriorated
dramatically, leaving a mean residual of $\pm$0$^{\prime\prime\!}$.99
and large systematic trends.  The conclusion was that the companion's
motion could not be fitted without a nongravitational acceleration.
Good news for the dirty snowball!

Three, the successful five-parameter solution fitting the data over a period
of more than two years allowed me to predict, with some confidence, the
companion's positions in Roemer's two exposures from 25~September 1960
and subsequently to locate it under a magnifying glass on at least one plate.
According to Roemer (1960, 1962) the companion was too faint to detect in
1960; its image was never measured and detection never reported.

\begin{table}[b]
\vspace{0.7cm}  % 0.8
\hspace{-0.2cm}
\centerline{
\scalebox{1}{
\includegraphics{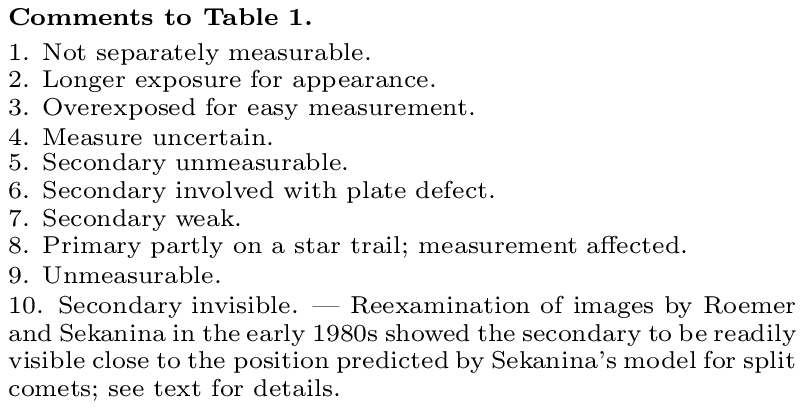}}} %  COMMENTS TO TABLE 1
\vspace{0.15cm} % 0.05
\end{table}

As for Roemer's unpublished results of her 1956--1960 astrometric
observations of comet Wirtanen's principal nucleus, they may exist
in her Putnam collection; if not, the relevant plates do.  For
related issues the reader is referred to Appendix B.

\begin{table}[t]
\vspace{0.11cm}
\hspace{-0.2cm}
\centerline{
\scalebox{0.96}{
\includegraphics{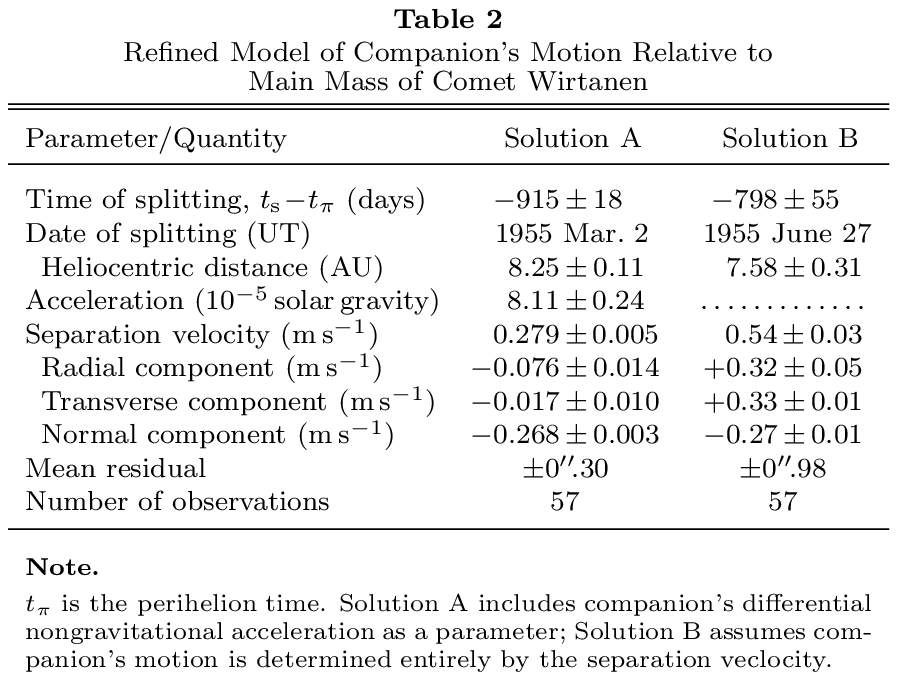}}} % TABLE 2
\vspace{0.7cm}
\end{table}

The companion's separation from the primary nucleus (relative positions)
is a different story altogether.  A paper copy of Roemer's measurement
records was recently found in an old folder of comet Wirtanen memorabilia
in the author's possession and the extensive data, converted to equinox
J2000, are listed~in~\mbox{Table}~1.  Even~though~the sender and
time of receipt are unknown, the document probably did not arrive until
after the work on the new model of the split comets (cf.\ Section~3) had
been completed.  The dataset in the document, effectively under embargo, was
apparently never used in any study other than Roemer's (1962, 1963).  Only
because of the unusual circumstances is it finally possible to disseminate
these valuable observations --- more than 60~years after they were made.

Table 1 indicates that even though Roemer took 58~exposures of comet
Wirtanen at the U.S.\ Naval Observatory's Flagstaff Station between
April~1957 and September~1960, both fragments were  actually measured
on only 48 of them.  Next, it was possible to combine these with 12 such 
measurements by the Lick Observatory's team (Jeffers \& Klemola 1958;
Jeffers \& Gibson 1960) and~with 19 by Van Biesbroeck (1961) to refine
the fragmentation solution published by Sekanina (1978) and noted above.
Close inspection of these 79 data points showed that at a rejection cutoff
as low as 0$^{\prime\prime\!}$.8 in either equatorial coordinate, 57 could
be retained for the least-squares solution:\ 41 by Roemer (85~percent of
her measured total), 9 by Jeffers et al.\ (75~percent), and 7 by Van
Biesbroeck (37~percent of his total; 55~percent{\vspace{-0.04cm}} with the
208-cm reflector but only 12~percent with the 61-cm reflector).\footnote{The
companion's position angles on \mbox{23--26}~August 1957 were given by
Van Biesbroeck incorrectly as 329$^\circ$, 330$^\circ$, and 331$^\circ$,
respectively.  Since the companion was to the west of the south (rather
than to the west of the north) of the primary, the position angles were
read as complements of 540$^\circ$, making the correct values equal to
211$^\circ$, 210$^\circ$, and 209$^\circ$, respectively.  The three
positions then passed as accurate enough for use in the solution.}

The five-parameter solution (Solution A) is compared in Table~2 with
the four-parameter solution (Solution~B), which is based on the same
57~observations but ignores the nongravitational acceleration.  Not
only is the mean residual of Solution~B more than a factor of 3
higher, but strong systematic trends, reaching 3$^{\prime\prime}$,
are distinctly apparent in the distribution of the residuals in Table~3.
Solution~A is a much refined version of the equivalent solution in
Sekanina (1978).  It suggests that the comet split at a heliocentric
distance exceeding 8~AU and confirms the veracity of the conclusion
that the effects of the nongravitational force on the companion's
motion were indeed robust.

The adopted rejection cutoff of 0$^{\prime\prime\!}$.8 is unusually tight,
yet it was passed by nearly three times as many observations as were
included in the 1978 dataset, whose rejection cutoff was much higher
than 1$^{\prime\prime}$.  The parameters of alternative solutions that
would be based on datasets with rejection cutoffs higher than
0$^{\prime\prime\!}$.8 would be similar to those in Table~2, but their
formal errors would naturally be higher.  We found that for the rejection
cutoff of 1$^{\prime\prime}$ the dataset would be augmented from 57 to
only 60 (all additional observations by Roemer), while for a
cutoff of 1$^{\prime\prime\!}$.5, the dataset would expand to 69 (four
additional observations by Roemer, three by the Lick Observatory's team,
and two by Van Biesbroeck).  This shows that relaxing the rejection cutoff
within reasonable limits would not enlarge the dataset's size very much.

\begin{table*}
\vspace{-0.05cm}
\hspace{-0.2cm}
\centerline{
\scalebox{1}{
\includegraphics{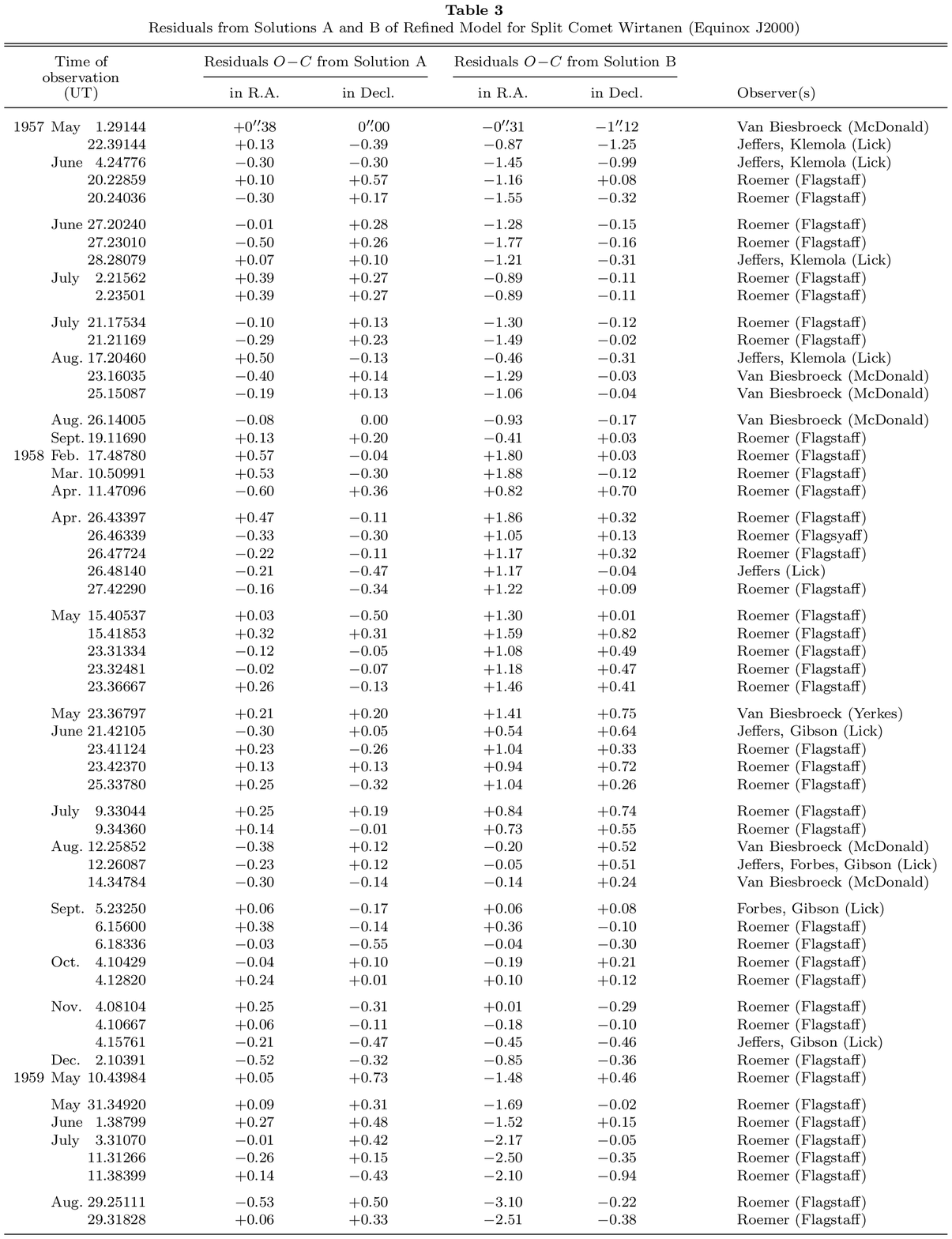}}}   % TABLE 3
\vspace{0cm}
\end{table*}

The companion's trajectory referred to the primary is plotted in Figure~1.
The upper part exhibits the 57~observations and their least-squares fit
by Solution~A (the five-parameter solution), the lower part displays the
trajectories derived, respectively, from Solutions~A and B, without the
observations.  The excellent fit by Solution~A is no surprise, merely
illustrating the very small residuals apparent from Table~3.

Of greater interest is the difference between the two solutions.
Solution~B appears to be ``bouncing'' around Solution~A over much of
1957 and stays north of it during the first half of 1958.  The inferior
fit by Solution~B is most strikingly documented by its divergence
from the fit by Solution~A around mid-1959, when the absence~of the
nongravitational acceleration is responsible for the failure to accommodate
the companion's continuing westward motion, taking instead a northward
path.  The net result is that the size of the gap between Solutions~A
and B more than doubles over the period of time between the second half
of 1959 and the second half of 1960.

By predicting the companion's position in September 1960, the potential
of the fragmentation model was exploited to its full advantage.  The
efforts to detect and measure the companion's images on Roemer's 1960
plates became the next focus of this investigation.

\begin{figure*}
\vspace{0.18cm}
\hspace{-0.2cm}
\centerline{
\scalebox{0.84}{
\includegraphics{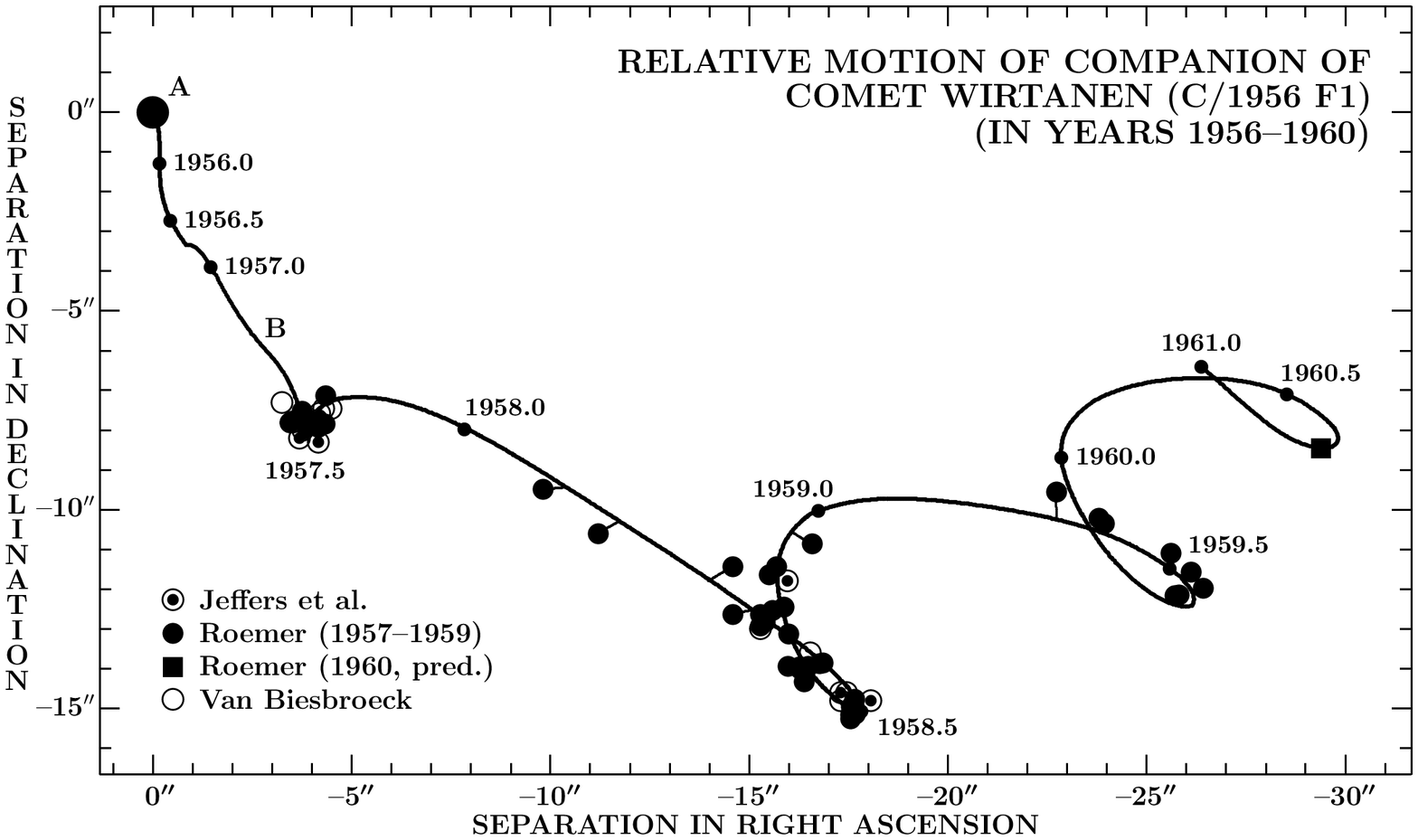}}} %  FIGURE 1

\vspace{0.9cm}
\hspace{-0.2cm}
\centerline{
\scalebox{0.84}{
\includegraphics{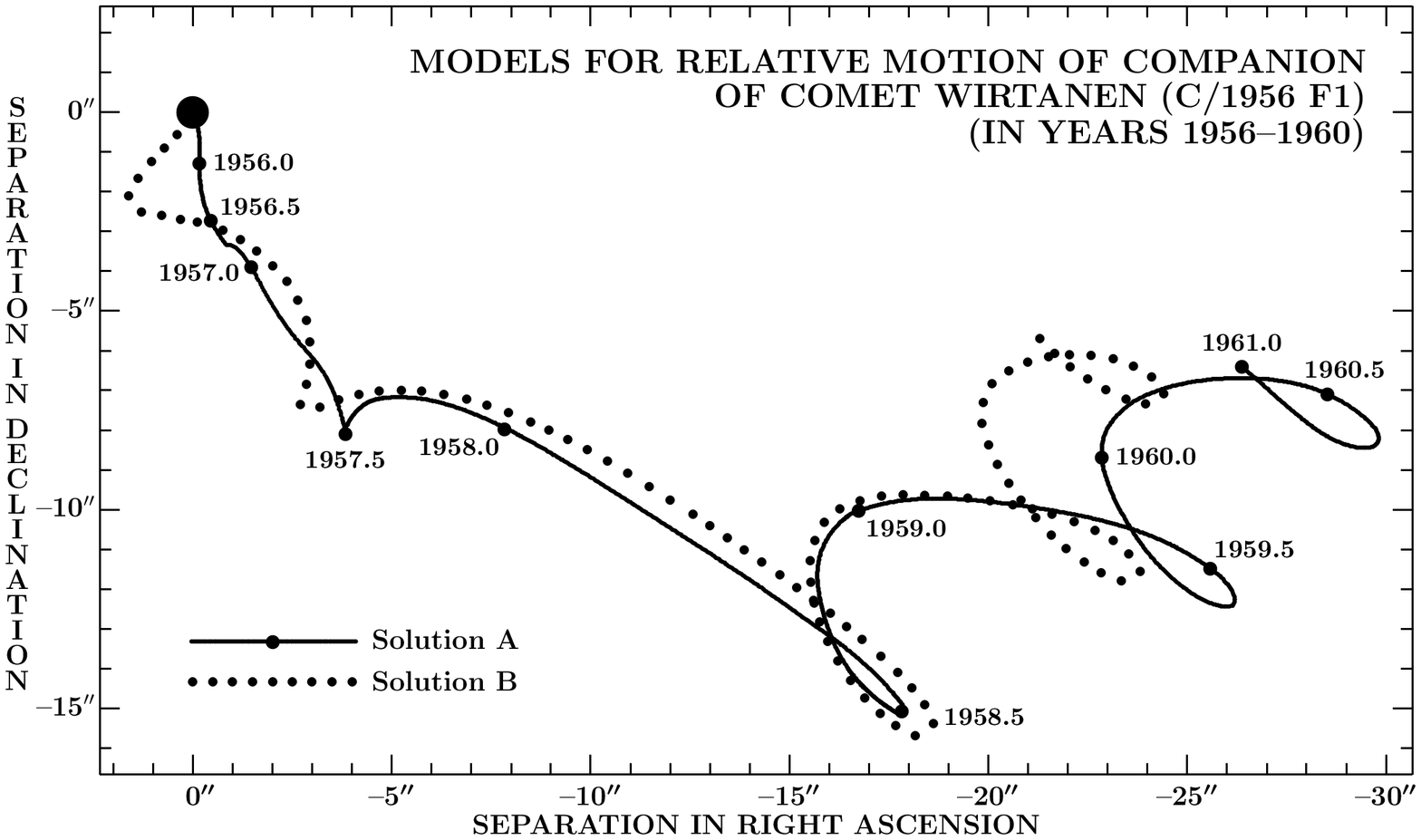}}}
\vspace{0cm}
\caption{Model of the motion of companion B relative to the primary A of
comet Wirtanen.  {\it Upper panel\/}:\ Comparison of Solution~A (which
includes nongravitational acceleration as parameter) from Table~2 with
57 observations that left residuals smaller than 0$^{\prime\prime\!}$.8
in either coordinate.  The square is the predicted position of the
companion on the single pair of 1960 plates taken by Roemer in
September.~{\it Lower panel\/}:\ Comparison of Solution~A (solid curve)
with the optimized four-parameter Solution~B (dotted curve).  See the
text for more comments.}
\end{figure*}

\section{Detection and Positions of Companion\\on the 1960 Plates}
% Section 5

As shown in Table 1, Roemer observed~comet~\mbox{Wirtanen} in 1960
on only a single night, 25~September,~when~she took two successive
exposures, ER 1069 and 1070.  These were the sole observations of
the comet made worldwide in 1960 as well as its final observations.
The~\mbox{companion} was predicted by the refined model (Solution~A)
to~be~located 30$^{\prime\prime\!}$.6 from the primary in a position
angle~of~253$^\circ\!$.9 (eq.\ J2000), which is equivalent to
the~offsets~of~$-$29$^{\prime\prime\!}$.4~in right ascension and
$-$8$^{\prime\prime\!}$.5 in declination (Figure~1).  The comet
was 1119~days past perihelion, 9.43~AU from the Sun, 8.65~AU from
the Earth, and located in the constellation of Pegasus, 24$^\circ\!$.5
to the north of the celestial equator.  The solar elongation of the
comet was 139$^\circ$ and its phase angle~4$^\circ\!$.

\begin{table}[t]
\vspace{0.18cm}
\hspace{-0.21cm}
\centerline{
\scalebox{1}{
\includegraphics{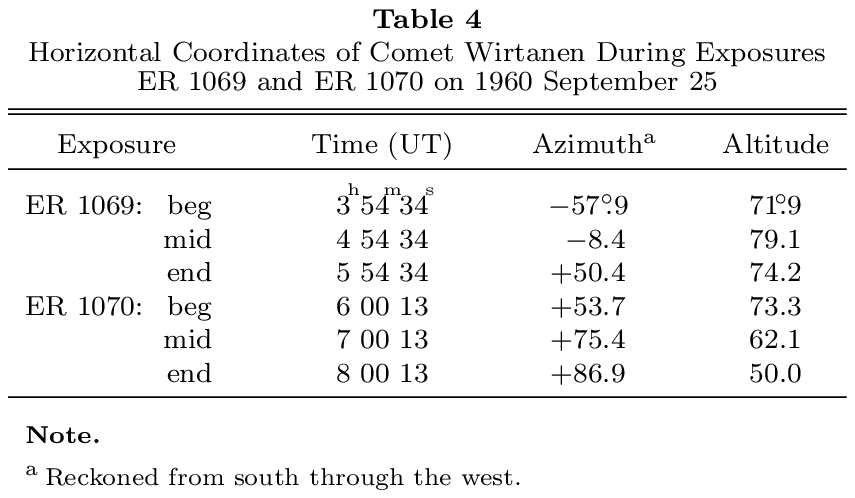}}} %  TABLE 4
\vspace{0.7cm}
\end{table}

The 102-cm f/6.8 Ritchey-Chr\'etien reflector of the U.S.\ Naval
Observatory's Flagstaff Station (code 689) was operated at an elevation
of 2310~meters above sea level and photographic images had a scale of
almost exactly 30$^{\prime\prime}$ per mm.  The two final exposures of
comet Wirtanen taken by Roemer were 120~minutes each, the emulsion was
Kodak 103a-O, and their mid-exposure times were recorded by the observer
% ; the companion's predicted
% distance from the main mass on the 1960 plates was a little over 1~mm.
% The emulsion was Kodak 103a-O.
% The mid-exposure times of the two 120-minute obser\-vations were recorded
% by Roemer 
as 1960 September 25.20456~UT for ER~1069 and 25.29182~UT
for ER~1070.  The comet was moving at a rate of 11$^{\prime\prime\!}$.4
per hour in position angle of 220$^\circ\!$.1.  The Moon was four
days old and the numbers show that the first exposure started just
as the Moon was setting.  The horizontal coordinates of the comet
during the exposures in Table~4 indicate that its zenithal distances
were always smaller than 19$^\circ$ during the first exposure, but
up to 40$^\circ$ toward the west at the end of the second exposure.
While these conditions are deemed quite satisfactory, the two
exposures do differ in quality, as mentioned below.

\begin{figure}[b]
\vspace{0.9cm}
\hspace{-0.22cm}
\centerline{
\scalebox{1.1}{ % 1.27
\includegraphics{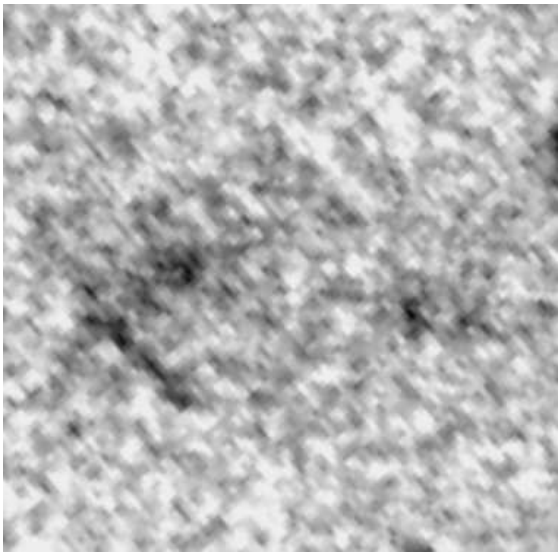}}} % FIGURE 2
\vspace{-0.05cm}	
\caption{Image of comet Wirtanen on plate ER~1069, exposed by
Roemer on 1960 September 25.2~UT.  The principal nucleus is on the
left, the companion on the right.  They are 31$^{\prime\prime}$ apart.
A faint star's trail is below
the principal nucleus.  The field extends about 105$^{\prime\prime}$
along the diagonal.  At the comet, 1$^{\prime\prime}$ is equivalent
to 6270~km.  North is up, east to the left.  The exposure time is
120~minutes, the emulsion is blue sensitive Kodak 103a-O.  The
plate scanned and processed by S.\ E.\ Levine, Lowell Observatory.
(From the Roemer collection at the Putnam Collection Center, Lowell
Observatory.){\vspace{-0.1cm}}}
\end{figure}

Following Roemer's death in April 2016, thousands of photographic plates
she obtained during her professional career have been archived in a
special collection at the {\it Putnam Collection Center\/} of the Lowell
Observatory in Flagstaff, Arizona.  The two 1960 plates were located~in
the collection by the staff of the Center and in response to my request
for assistance with their inspection B.~A.~Skiff arranged that the
plates were scanned and processed by S.\ E.\ Levine.  The scanning
was done at 2400~dpi on a desktop flatbed scanner at Lowell.  The pixel
size of 10.58~micron was equivalent to 0$^{\prime\prime\!}$.32 at the
plate resolution.  A region of about 10$^\prime \!\times\! 10^\prime$
centered on the comet's image on either plate was selected and digitally
stored in fits format.  A .png version of the file was rotated and the
contrast and brightness slightly adjusted by the author, who applied
the GIMP software package.  The generated .eps image on ER~1069 is
shown in Figure~2.

The field in the figure measures approximately 105$^{\prime\prime}$ along
the diagonal and is oriented with the north up and east to the left.  The
principal nucleus of comet Wirtanen is a little to the left and the
companion to the right of the center, while a faint star below the main
mass is trailed because the telescope was tracking the comet during the
two-hour exposure.  The companion was measured to be 31$^{\prime\prime}$
from the principal nucleus, equivalent to a projected distance of
$\sim$194,000~km, in a position angle of between 255$^\circ$ and
257$^\circ\!$.  The uncertainty is estimated at more~than
$\pm$1$^{\prime\prime}$ in the separation distance and more than
$\pm$2$^\circ$ in the position angle.

The second plate, ER 1070, was distinctly darker than the first, which is
somewhat surprising given that the comet was never more than 40$^\circ$
from the zenith.  The companion's image on this plate, inferior in quality
to ER~1069, is less obvious, although a clump of dark pixels is apparent
at the predicted location as well.

On the original plate ER 1069, the companion is discernible visually under
high magnification.  However, since the comet had not been observed for
more than a year before the final exposures, the companion's position
relative to the principal nucleus in late September 1960 was anybody's
guess.  In the absence of a sound fragmentation model, the companion's
position could not be {\it predicted\/}, only {\it inferred\/} on certain
assumptions from the separation curve before 1960.  For example, straightforward
extrapolation of the companion's motion from Roemer's observations during
1959 suggested that at the time the 1960 observation the object should
have been located some 45$^{\prime\prime}$ from the principal nucleus in
a position angle near 236$^\circ$, that is, nearly 20$^{\prime\prime}$
(sic!) to the southwest of the actual position; no object at all is
seen on the plate at that location.  I believe that it was this kind
of argument --- not questioning the existence of the feature at
31$^{\prime\prime}$ from the principal nucleus, a nonissue in the
1960s --- that was behind Roemer's conclusion that the companion was
no longer visible.  Conversely, the fact that the position of the
feature in Figure~2 is compatible with the position predicted by the
fragmentation model greatly strengthens the conclusion that the observed
feature indeed {\it is\/} the companion.

\section{Conclusions}
Attention is called to a large number of high-quality astrometric
observations of the main and companion nuclei of the Oort-Cloud object
Wirtanen, made by E.~Roemer at Lick in 1956 and at Flagstaff in
1957--1960, which were {\it never published\/}.  This is peculiar
both because she was a prolific observer who did report thousands of
astrometric positions of comets and unusual asteroids, and especially
hard to comprehend in the light of her statement that Wirtanen was an
exceptionally interesting comet.

From the standpoint of investigations of cometary fragmentation, the
positions of companions relative to the main nucleus are of particular
interest.  Recently, a two-page document in Roemer's handwriting,
providing a set of relative positions of the companion to the principal
nucleus of comet Wirtanen (48~separation distances and position angles)
was fortuitously discovered in my old folder on comet Wirtanen.  While
I have no recollection of how the document got in, its contents are
replicated in Table~1, with the position angles duly converted from
Roemer's equinox 1900 to J2000.  

Combined with the observations reported by the Lick team and by Van
Biesbroeck, a set of 57~highly consistent data is now employed to refine
the parametric values of the fragmentation model (Solution~A in Table~2).
Even though the model was developed some 45~years ago (Sekanina 1978),
its application to comet Wirtanen was then limited to only an inferior
set of 22~data, as Roemer's observations were of course unavailable.
The versatility of the model is once again exploited to confirm that
the companion's motion in the orbital plane was governed by the differential
nongravitational acceleration and that the ``traditional'' solution based
on the assumption that the companion's motion was controled instead by the
separation velocity (Solution~B in Table~2) failed miserably.

The parameters of the presented fragmentation model for comet Wirtanen
(Solution~A) --- the companion's separation in early March 1955, the
velocity of separation of only 0.28~m~s$^{-1}$, mostly in the out-of-plane
direction, and the nongravitational acceleration of 8.1 units of
10$^{-5}$ the solar gravitational acceleration --- differ dramatically
from the results of Roemer's (1962, 1963) model, which ignored not only
the effects of the acceleration, but likewise the temporal variations in
the position angle of the companion's motion relative to the principal
nucleus.

The refined fragmentation model has also been used to predict the
companion's position at the time of the final observation of the comet
on 1960 September~25, when according to Roemer's (1960) report the
companion was no longer visible.  Although this claim was previously
disputed based on brief (and unrecorded) inspection~of the 1960
images, a more formal and comprehensive~inves\-ti\-gation was
desirable.  In collaboration with my~Lowell Observatory's colleagues,
such an examination has~now been completed with the result that the
companion has been detected at a position that agrees, within the
errors of measurement, with the position predicted by the model.
The images of the principal nucleus and the companion on one of
the plates are presented.\\[0.35cm]

{\it Acknowledgements.}$\!$ --- I am thankful to Stephen E.\ Levine, Lowell
Observatory, for scanning and processing the plates ER~1069 and ER~1070 of
comet Wirtanen; Brian A.\ Skiff, Lowell Observatory, for his kind response
to my request for assistence with the project and a plate search; and Lauren
Amundson, the Putnam Collection Center, Lowell Observatory, for making the
plates available and for information on the Roemer Collection.  I also thank
Daniel W.\ E.\ Green, Director of the Central Bureau for Astronomical
Telegrams, Department of Earth and Planetary Sciences, Harvard University,
for providing me with a set of requested older IAU Circulars.

\begin{center}
{\large \bf Appendix A}\\[0.3cm]
COMMENTS ON ROEMER'S INVESTIGATION OF\\THE SPLITTING OF COMET WIRTANEN
\end{center}
With the companion's separation from the primary, measured by Roemer from
her plates, now available in Table~1, it should be possible to replicate
the procedure and the conclusions she arrived at by plotting the linear
distance of the two fragments projected onto the plane of the sky as a
function of time (Roemer 1962, 1963).  Merely copying her approach would
not be of much interest, if the conclusions did not differ from hers.

\begin{figure}[b]
\vspace{0.8cm}
\hspace{-0.7cm}
\centerline{
\scalebox{0.505}{
\includegraphics{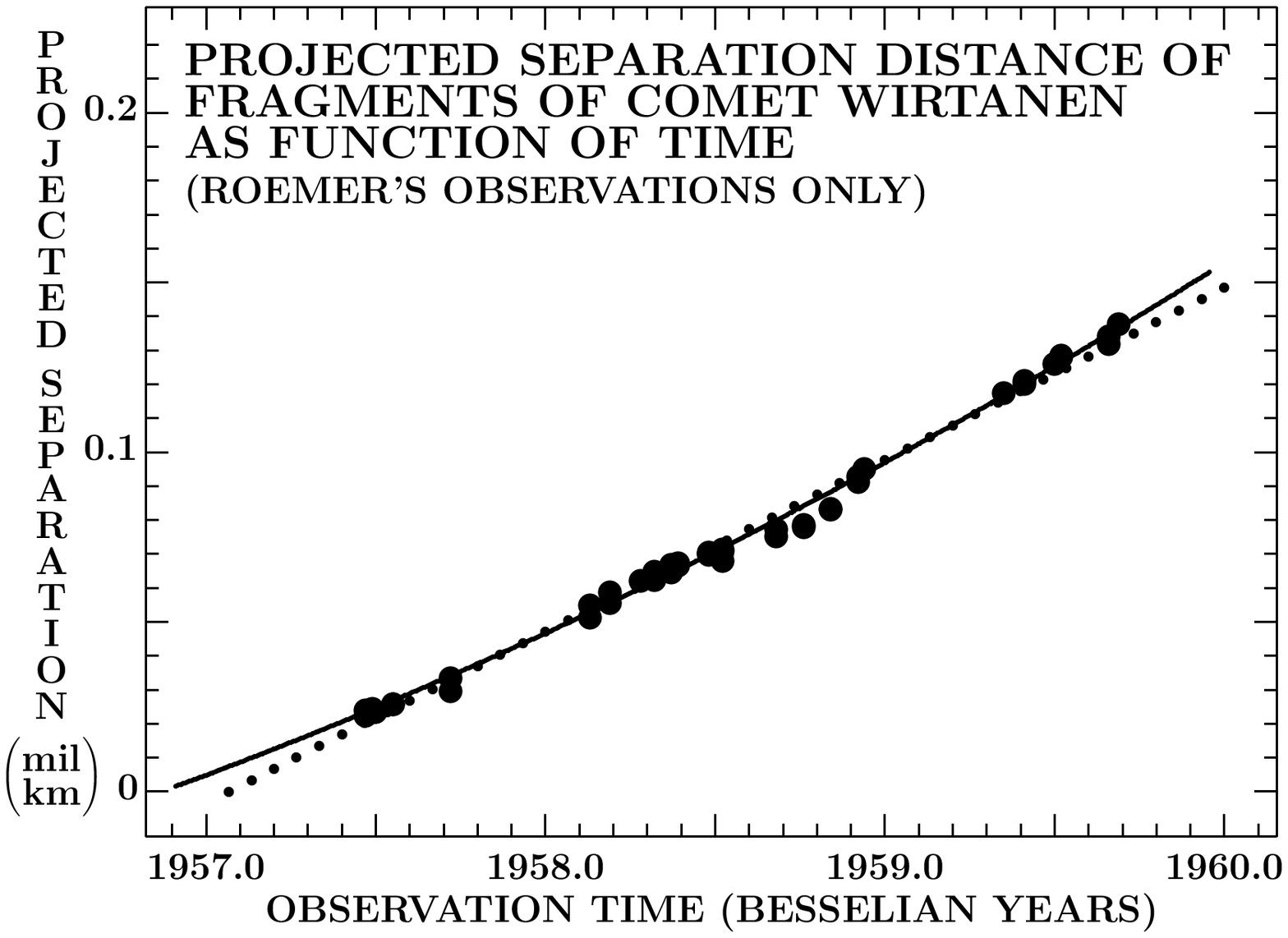}}} % FIGURE A-1
{\footnotesize {\bf Figure A-1.}  Plot of the projected separation distance
between the two nuclei of comet Wirtanen as a function of time.  All 48
data points are Roemer's observations.  The dotted line is a linear
least-squares fit, the solid curve a quadratic least-squares
fit.{\vspace{-0.02cm}}}  
\end{figure}

The increasing projected separation between the two fragments is
displayed in Figure~A-1.  Plotted are only Roemer's measurements, so
that I use the same dataset as she did.~Expressing time $t$ in the
Besselian~years~(each equaling \mbox{365.2422 days = $3.1557 \:\!\!
\times \!10^7$\,s}) and the separation distance $\Delta_{\rm sep}$
in millions of kilometers, I obtain by least squares from Roemer's
48~observations:
\begin{eqnarray*}
\hspace{0.4cm} \Delta_{\rm sep} & = & -0.0037 + 0.0506\,\tau,
  \hspace{3.27cm}\mbox{(A-1)}\\[-0.05cm]
  & & \pm 0.0012 \hspace{0.05cm} \pm \!0.0007  \nonumber % (A-1)	     
\end{eqnarray*}
where \mbox{$\tau = t - B1957.0$}.  In 1957 the Besselian year~\mbox{began}
on January~0.619~TT, so that 1957~Jan.~1.0 TT equals B1957.00104.  Defining
--- as Roemer did --- the time of splitting, $t_{\rm s}$, by
\mbox{$\Delta_{\rm sep} = 0$}, I get from Equation~(A-1)
\begin{displaymath}
\hspace{0.4cm} t_{\rm s} = B1957.073 \pm 0.024 = 1957 \; {\rm Jan.} \; 26
	\pm 9 \; {\rm TT}. \hspace{0.15cm}\mbox{(A-2)} % (A-2)
\end{displaymath}
Roemer (1962, 1963) wrote that the breakup took place at the beginning
of January 1957 with an uncertainty of a few days.  The projected
velocity of separation, $U_{\rm sep}$, defined by the slope
of the curve (dotted line in the plot), equals 0.0506 million km per
year or
\begin{displaymath}
\hspace{0.4cm} U_{\rm sep} = \dot{\Delta}_{\rm sep} = 1.60 \pm 0.02 \;
  {\rm m} \; {\rm s}^{-1}\!, \hspace{2.34cm}\mbox{(A-3)} %  (A-3)
\end{displaymath}
close to Roemer's value of 1.5~m~s$^{-1}$.  The somewhat later fragmentation
time above is inevitably the consequence of the slightly higher velocity.

The most important property of Figure~A-1 was~\mbox{either} missed or
deemed questionable by Roemer to remark~on:\ the relationship {\it is
not perfectly linear\/}, the separation distance increasing with time
at a slightly higher power.  This deviation from the linearity is
readily accounted for by assuming a quadratic dependence.  A
least-squares fit offers in this case, in the same units as in
Equation~(A-1), the following expression
\begin{eqnarray*}
\hspace{0.4cm} \Delta_{\rm sep} & = & 0.0045 + 0.0376 \, \tau + 0.00425
  \, \tau^2.\hspace{1.6cm}\mbox{(A-4)} \\[-0.05cm] % (A-4)
  & & \hspace{-0.28cm}\pm 0.0007 \hspace{0.05cm}\pm \! 0.0024
      \hspace{0.31cm}\pm \! 0.00075 \nonumber
\end{eqnarray*}
When \mbox{$\Delta_{\rm sep} = 0$}, the relevant time of splitting is
the root of this quadratic equation with the positive sign in front of
the square root of the discriminant:
\begin{displaymath}
\hspace{0.36cm} t_{\rm s} = B1956.879 \pm 0.030 = 1956 \: {\rm Nov.}
	\, 17 \pm 11 \, {\rm TT}. \hspace{0.13cm} \mbox{(A-5)}  %  (A-5)
\end{displaymath}
The projected relative velocity, expressed in millions of km per year,
now equals
\begin{eqnarray*}
\hspace{0.4cm} \dot{\Delta}_{\rm sep} & = & 0.0376 + 0.0085 \, \tau.
  \hspace{3.61cm} \mbox{(A-6)} \\[-0.05cm] % (A-6)
  & & \hspace{-0.28cm} \pm 0.0024 \hspace{0.05cm} \pm \! 0.0015 \nonumber
\end{eqnarray*}
Inserting $t_{\rm s}$ I get for the separation velocity, after conversion
to the preferred units,
\begin{displaymath}
\hspace{0.4cm} U_{\rm sep} = 1.16 \pm 0.08 \; {\rm m} \; {\rm s}^{-1}.
  \hspace{3.48cm} \mbox{(A-7)} %  (A-7)
\end{displaymath}
The second term on the right-hand side of Equation\,\mbox{(A-6)} indicates
that the projected relative velocity was increasing at an average rate
of 0.0085 million km per year per year, or
\begin{displaymath}
\hspace{0.4cm} \ddot{\Delta}_{\rm sep} = (0.085 \pm 0.015) \:\!\!
  \times \! 10^{-5} \: {\rm cm} \; {\rm s}^{-2}. \hspace{1.7cm}
  \mbox{(A-8)} % (A-8)
\end{displaymath}

In order to be able to judge the magnitude of this acceleration, it is
first necessary to normalize it.  With the previously employed inverse
square power law, the equivalent average projected rate of
acceleration at 1~AU from the Sun is obtained by multiplying the value
from the expression{\vspace{-0.07cm}} (A-8) by a factor of
$\langle r^2 \rangle/r_{\!\mbox{\tiny \boldmath $\oplus$}}^2$, where
$\langle r^2 \rangle$ is the average value of $r^2$ for the 48
observations by Roemer and \mbox{$r_{\!\mbox{\tiny \boldmath $\oplus$}}
= 1$ AU}.  Numerically the factor equals 28.9, so that the {\it normalized\/}
projected acceleration between the motions of the two fragments comes
out to be
\begin{eqnarray*}
\hspace{0.4cm} \langle \gamma \rangle & = & (2.5 \pm 0.4) \:\!\! \times
	\! 10^{-5} \: {\rm cm} \; {\rm s}^{-2} \\[-0.05cm]
    & = & (4.2 \pm 0.7) \:\!\! \times \! 10^{-5} \:{\rm solar \,\,
	grav. \: acceleration.} \hspace{0.22cm} \mbox{(A-9)} % (A-9)
\end{eqnarray*}
I recall that the total acceleration, $\gamma$, was determined to equal
\mbox{$8.1 \:\!\! \times \! 10^{-5}$ the} solar gravitational acceleration,
nearly twice higher than $\langle \gamma \rangle$, from{\vspace{-0.03cm}}
the much more rigorous approach described in Section~4 (see
Table~2).\footnote{The acceleration's values derived from the total
of 57~observations on the one hand, and from Roemer's observations
alone on the other hand, are practically identical.}

One can go a step further and examine the magnitude of the projected
acceleration on{\vspace{-0.05cm}} certain assumptions.  If {\boldmath
$\ddot{\Delta}$}$_{\rm sep}(t)$ is a unit acceleration vector and
{\boldmath $\Delta$}$_{\rm earth}(t)$ a unit geocentric-distance vector of
the comet at time $t$, the subtended angle $\theta(t)$ measures the degree
of foreshortening of the acceleration vector in projection onto the plane
of the sky, given by the vector product
\begin{displaymath}
\hspace{0.4cm} \sin \theta(t) = \left| \mbox{\boldmath
  $\ddot{\Delta}$}_{\rm sep}(t) \mbox{\large \boldmath $\times$}
  \mbox{\boldmath $\Delta$}_{\rm earth}(t) \right|. \hspace{2.4cm}
  \mbox{(A-10)} % (A-10)
\end{displaymath}
Over a period of time, the average projected magnitude of the acceleration,
$\langle \gamma \rangle$, is related to its total magnitude, $\gamma$, by
\begin{displaymath}
\hspace{0.4cm} \gamma \, \langle \sin \theta \rangle = \langle \gamma
  \rangle, \hspace{4.95cm} \mbox{(A-11)} % (A-11)
\end{displaymath}
where $\langle \sin \theta \rangle$ is an average value of $\sin \theta$
over the time period.  For example, if the acceleration vector should
relatively to the Earth direction be distributed nearly at random,
the average would be \mbox{$\langle \sin \theta \rangle = \frac{1}{2}$},
{\vspace{-0.06cm}}so that \mbox{$\langle \gamma \rangle =
4.05 \:\!\! \times \! 10^{-5}$\,the} solar gravitational acceleration,~a
value close to that in Equation~(A-9).  It appears that unlike for the
time of splitting and the separation velocity, Roemer's plot offers a
meaningful value for the companion's differential nongravitational
acceleration.\\
\begin{center}
{\large \bf Appendix B} \\[0.3cm]
ROEMER'S ASTROMETRIC POSITIONS OF\\THE PRINCIPAL NUCLEUS OF\\COMET
 WIRTANEN
\end{center}
This appendix provides information that as of now (early 2023) I have
been able to ascertain on astrometric observations of comet Wirtanen
made by Roemer at the Lick Observatory in 1956 and at the U.S. Naval
Observatory's Flagstaff Station in 1957--1960.  My effort has focused
on two issues:\ (i) Did she at any time determine the astrometric
positions of the main mass? (ii) If yes, is there any evidence that
she either published or otherwise disseminated these results?

To address the two issues, I have carefully inspected relevant
publications for indications that Roemer's astrometric positions of
the principal nucleus were employed one way or the other, especially
to compute the orbital elements and/or ephemeris, whether by her or
others.  I found that at least some of her observations, made on
selected dates, were indeed used.  For example, Mowbray (1956) reported
his new orbit determination {\it ``based on observations \ldots by
Dr.~Elizabeth Roemer on March~20, April~2, and April~30,''} 1956.
These observations were made by her at the Lick Observatory.  Similarly,
Roemer (1958) herself employed her Flagstaff observations from
16~September 1957 (plates ER~118 and 119) and 26~April 1958 (plates
ER~289--291) to determine the corrections to Mowbray's orbit.

While inspecting the old literature, I found --- with much surprise
and chagrin --- that I had used ten observations made by Roemer at
Lick between 20~March and 28~May 1956 to compute the orbit in my
early work on comet Wirtanen (Sekanina 1968).  The paper listed the
individual residuals --- observed minus calculated from the orbit
--- but not the observed positions.  All entries by Roemer were
referenced as a private communication from her in 1967.  She obviously
sent them upon request and it is likely that their use was conditioned
on stipulation that the observations not be listed in the paper, even
though it was straightforward for the reader to derive them by adding
the residuals to the coordinates computed from the orbit.  None of the
observations made by Roemer in the years 1957 through 1960 were employed
in the paper.

To summarize, I am now ready to address the two questions asked at
the beginning of this appendix.  It is certain that ten observations
of the comet's principal nucleus made by Roemer at Lick in March--May
1956 were astrometrically reduced.  It also appears that at least some
--- and possibly all --- observations of this primary nucleus she made at
the U.S.\ Naval Observatory's Flagstaff Station in 1957--1960 were also
fully reduced.  At least a fraction of these data were made available by
Roemer via personal communications to others, apparently upon request, in
support of orbit-determination efforts.  Nonetheless, there is absolutely
no evidence that even a single of these astrometric positions has ever
been published.  The numbers suggest that the dataset may contain up to
about 70~entries and should be located in Roemer's collection in the {\it
Putnam Collection Center\/} of the Lowell Observatory.  An effort aimed
at finding these astrometric positions of the principal nucleus is
outside the scope of this investigation, but such a search is strongly
recommended because, if successful, it should facilitate a major
refinement of the orbit of comet Wirtanen in the future.\\[-0.36cm]
\begin{center}
{\footnotesize REFERENCES}
\end{center}
\vspace{-0.4cm}
\begin{description}
{\footnotesize
\item[\hspace{-0.3cm}]
Jeffers, H.\ M., \& Gibson, J.\ 1960, AJ, 65, 163
\\[-0.57cm]
\item[\hspace{-0.3cm}]
Jeffers, H.\ M., \& Klemola, A.\ R.\ 1958, AJ, 63, 249
\\[-0.57cm]
%
% \item[\hspace{-0.3cm}] 
% Marsden, B.\ G. 1969, AJ, 74, 720
% \\[-0.57cm]
%
% \item[\hspace{-0.3cm}]
% Marsden, B.\ G. 1970, AJ, 75, 75
% \\[-0.57cm]
%
% \item[\hspace{-0.3cm}]
% Marsden, B.\ G., \& Sekanina, Z. 1974, AJ, 79, 413
% \\[-0.57cm]
%
\item[\hspace{-0.3cm}]
Marsden, B.\ G., Sekanina, Z., \& Everhart, E.\ 1978, AJ, 83, 64
\\[-0.57cm]
\item[\hspace{-0.3cm}]
Marsden, B.\ G., Sekanina, Z., \& Yeomans, D.\ K.\ 1973, AJ,~78,~211
\\[-0.57cm]
\item[\hspace{-0.3cm}]
Mowbray, A.\ G.\ 1956, IAU Circ.\ 1555
\\[-0.57cm]
\item[\hspace{-0.3cm}]
Roemer, E.\ 1958, IAU Circ.\ 1638 \& 1647
\\[-0.57cm]
\item[\hspace{-0.3cm}]
Roemer, E.\ 1960, PASP, 72, 512
\\[-0.57cm]
\item[\hspace{-0.3cm}]
Roemer, E.\ 1962, PASP, 74, 351
\\[-0.57cm]
\item[\hspace{-0.3cm}]
Roemer, E.\ 1963, AJ, 68, 544
\\[-0.57cm]
\item[\hspace{-0.3cm}]
Roemer, E.\ 1965, AJ, 70, 397
\\[-0.57cm]
\item[\hspace{-0.3cm}]
Roemer, E., \& Lloyd, R.\ E.\ 1966, AJ, 71, 443
\\[-0.57cm]
\item[\hspace{-0.3cm}]
Roemer, E., Thomas, M., \& Lloyd, R.\ E.\ 1966, AJ, 71, 591
\\[-0.57cm]
\item[\hspace{-0.3cm}]
Sekanina, Z.\ 1968, Bull.\ Astron.\ Inst.\ Czech., 19, 153
\\[-0.57cm]
\item[\hspace{-0.3cm}]
Sekanina, Z.\ 1975, Icarus, 25, 218
\\[-0.57cm]
\item[\hspace{-0.3cm}]
Sekanina, Z.\ 1977, Icarus, 30, 574
\\[-0.57cm]
\item[\hspace{-0.3cm}]
Sekanina, Z.\ 1978, Icarus, 33, 173
\\[-0.57cm]
\item[\hspace{-0.3cm}]
Stefanik, R.\ P.\ 1966, M\'em.\ Soc.\ Roy.\ Sci.\ Li\`ege
 (S\'er.\ 5), 12, 29
\\[-0.57cm]
\item[\hspace{-0.3cm}]
Van Biesbroeck, G.\ 1961, AJ, 66, 96
\\[-0.63cm]
\item[\hspace{-0.3cm}]
Whipple, F.\ L.\ 1950, ApJ, 111, 375}
% \\[-0.57cm]
%
\vspace{-0.39cm}
\end{description}
\end{document}